\documentclass[11pt]{article}

\usepackage[margin=1in]{geometry}
\usepackage{amsmath,amssymb,bm}
\usepackage{graphicx}
\usepackage{booktabs}
\usepackage{array}
\usepackage{tabularx}
\usepackage{caption}
\usepackage{tikz}
\usetikzlibrary{arrows.meta,positioning,shapes.geometric,calc}
\usepackage[colorlinks=true,linkcolor=blue,citecolor=blue,urlcolor=blue]{hyperref}

\newcolumntype{L}[1]{>{\raggedright\arraybackslash}p{#1}}
\usepackage[english]{babel}
\usepackage[final]{microtype}

\title{\bfseries Physics-Guided Machine Learning for Predictive Turbulent Combustion Simulation}

\author{%
\textbf{Shubhangi Bansude}$^{1,*}$ and \textbf{Jay Patel}$^{1}$\\[0.6em]
$^{1}$Department of Mechanical Engineering, Indian Institute of Technology Gandhinagar,\\
Palaj, Gandhinagar, Gujarat, India\\[0.6em]
$^{*}$Corresponding Author: \texttt{shubhangi.bansude@iitgn.ac.in}
}

\date{}

\begin{document}
\sloppy
\maketitle

\begin{abstract}
\noindent
Predictive simulation of turbulent combustion remains challenging because stiff chemical kinetics, turbulent transport, molecular mixing, and heat release interact nonlinearly across unresolved scales.  In the filtered or averaged formulations, reaction rate is unclosed, and the integration of detailed chemistry constitutes the dominant computational cost. Conventional closures such as mixing-limited models, flamelet and manifold methods, conditional moment closure, and transported probability/filtered density function (PDF/FDF) formulations, encode substantial physical insight, but rely on structural assumptions that may lose validity outside their calibrated regimes. Machine learning offers a complementary approach because of its ability to approximate complex nonlinear functions. In turbulent combustion, this capability can be used to relax restrictive closure assumptions, learn unresolved nonlinear mappings from data, and accelerate expensive computations such as detailed chemistry computation. However, purely data-driven models may violate conservation laws and thermochemical consistency. They may also extrapolate poorly outside the training domain and destabilize the CFD solvers in which they are embedded. Physics-guided machine learning (PGML) addresses these failure modes by incorporating prior knowledge throughout the modeling pipeline: in the training data and input features, the model architecture, the loss function, the hybrid closure structure, and the solver-aware validation protocol. This article reviews PGML for turbulence-chemistry interaction (TCI) modeling within this framework. It formulates the learning problem, emphasizes the distinction between a priori and a posteriori assessment, and organizes the literature on chemistry acceleration, reduced order and manifold modeling, subgrid closure, and filtered density function closure, with soft versus hard constraint enforcement treated in detail. The author's research on neural ordinary differential equations (neural ODEs) for stiff chemical kinetics and on deep-learning filtered density function closure is discussed within the corresponding topics as illustrative applications of these principles. Finally, generalization, interpretability, uncertainty quantification, data infrastructure, and solver integration are examined as key requirements for developing accurate, robust, and CFD-ready combustion models.
\end{abstract}

\vspace{0.5em}
\noindent{\sloppy\textbf{Keywords:} turbulent combustion; physics-guided machine learning; large-eddy simulation; turbulence-chemistry interaction; chemical kinetics; neural ordinary differential equations; filtered density function\par}

\section*{Nomenclature}
\begin{center}\small
\begin{tabular}{llll}
$\rho$ & mixture density & $u_i$ & velocity component \\
$p$ & pressure & $T$ & temperature \\
$Y_\alpha$ & mass fraction of species $\alpha$ & $h$ & specific enthalpy \\
$Z$ & mixture fraction & $c$ & reaction progress variable \\
$\Phi$ & equivalence ratio & $\bm{\Phi}$ & vector of thermochemical scalars \\
$\dot{\omega}_\alpha$ & chemical source term of species $\alpha$ & $\bm{S}(\bm{\Phi})$ & chemical source term vector \\
$D_\alpha$ & molecular diffusivity of species $\alpha$ & $V_{\alpha,j}$ & diffusion velocity of species $\alpha$ \\
$\sigma_{ij}$ & viscous stress tensor & $\lambda$ & thermal conductivity \\
$\Delta$ & LES filter width & $\chi$ & scalar dissipation rate \\
$\tau_{ij}^{\mathrm{sgs}}$ & subgrid scale stress tensor & $F(\bm{\psi})$ & filtered density function \\
$\bm{\theta}$ & trainable model parameters & $f_{\bm{\theta}}$ & neural-network model \\
$\mathcal{L}$ & training loss function & $\lambda_k$ & weight of $k$-th physics penalty \\
\end{tabular}
\end{center}

\section*{Abbreviations}
\begin{center}\small
\begin{tabular}{llll}
ANN/DNN & artificial/deep neural network & LES & large-eddy simulation \\
BNN & Bayesian neural network & ML & machine learning \\
SciML & scientific machine learning & PGML & physics-guided machine learning\\
CFD & computational fluid dynamics & MLP & multilayer perceptron \\
CMC & conditional moment closure & NODE & neural ordinary differential equation \\
CNN & convolutional neural network & OOD & out-of-distribution \\
CRNN & chemical reaction neural network & PaSR & partially stirred reactor \\
DNS & direct numerical simulation & PCA & principal component analysis \\
EDC & eddy dissipation concept & PDF/FDF & probability/filtered density function \\
FGM & flamelet-generated manifold & PINN & physics-informed neural network \\
FPV & flamelet/progress variable & PMSR & pairwise mixing stirred reactor \\
FSD & flame surface density & RANS & Reynolds-averaged Navier-Stokes \\
GPR & Gaussian process regression & SGS & subgrid scale \\
ISAT & in situ adaptive tabulation & TBNN & tensor basis neural network \\
TCI & turbulence-chemistry interaction & UQ & uncertainty quantification \\
\end{tabular}
\end{center}

\section{Introduction}
\label{sec:intro}

Turbulent combustion powers most of the world's transportation, propulsion, and energy conversion systems, and it will remain central to aviation, heavy transport, and industrial heat during the transition toward hydrogen, ammonia, and sustainable fuels. Designing cleaner and more efficient combustors, therefore, depends on the ability to predict, rather than merely reproduce, the behavior of turbulent flames across operating conditions and geometries. This predictive goal remains difficult. Turbulent reacting flows couple a continuum of eddy scales with chemical processes whose time scales span more than ten orders of magnitude, from nanosecond radical reactions to residence times of milliseconds or longer \cite{peters2000,pope2013}. Heat release feeds back on the flow through density and viscosity variations. Molecular mixing controls the composition entering the reaction zones, and the chemical source term is an exponentially nonlinear function of temperature. As a consequence, the statistical moments evolved by averaged or filtered transport equations are not closed: the mean or filtered reaction rate cannot be expressed in terms of mean or filtered quantities alone \cite{poinsot2011}.

Decades of research have established a hierarchy of turbulent combustion closures, including eddy dissipation models, laminar flamelet and flamelet generated manifold methods, partially stirred reactor (PaSR) formulations, conditional moment closure, and transported probability density function (PDF) and filtered density function (FDF) methods \cite{magnussen1977,peters1984,klimenko1999,pope1985,colucci1998}. Each model embodies a specific physical hypothesis about turbulence-chemistry interaction (TCI) and has contributed substantially to predictive combustion simulation. However, these approaches also face persistent limitations, including the stiffness and cost of finite-rate chemistry, restrictive fast chemistry or manifold assumptions, empirically tuned parameters, and closure hypotheses whose validity may deteriorate outside the regimes for which they were developed or calibrated \cite{haworth2010,lu2009}. High-fidelity simulation of practical combustors therefore requires careful expert judgment to balance physical fidelity, computational cost, and numerical robustness.

A complementary resource has matured alongside these models: data. Petascale direct numerical simulations (DNS) now resolve TCI in canonical configurations approaching conditions of engineering relevance \cite{chen2011}. Quantitative laser diagnostics provide multidimensional measurements of scalar and velocity fields, while canonical reactor, flamelet, and flame databases can be generated systematically at comparatively low cost. In parallel, machine learning (ML), particularly deep neural networks (DNNs), has emerged as a powerful tool for approximating high-dimensional nonlinear mappings. These capabilities have enabled ML to be used across the combustion modeling workflow, including chemical kinetics acceleration, flamelet table compression, subgrid scale (SGS) closure modeling, low-dimensional manifold discovery, and surrogate or digital twin development for design and optimization \cite{ihme2022,zhou2022,brunton2020}.

Early applications also revealed a recurring difficulty, one encountered previously in data-driven turbulence modeling \cite{duraisamy2019}: models trained without regard for the governing physics can violate mass conservation and element balance, produce unbounded or unrealizable predictions, extrapolate poorly outside their training manifold, and destabilize the flow solvers into which they are inserted. The consensus that has emerged, articulated for combustion by Ihme et al. \cite{ihme2022} and for scientific machine learning (SciML) broadly by Karniadakis et al. \cite{karniadakis2021}, is that predictive data-driven combustion models must be \emph{physics-guided}: prior knowledge should be embedded in the data and features, the model architecture, the training objective, the surrounding hybrid model structure, and the validation protocol.

This article reviews the field within that framework.  Section~\ref{sec:closure} presents the governing equations, the closure problem, and the conventional models that data-driven methods seek to accelerate, augment, or replace; the notation established there is used throughout. Section~\ref{sec:pgml} presents physics-guided machine learning (PGML) as a unified modeling framework. It formulates the learning problem, discusses appropriate assessment strategies, describes different mechanisms for embedding physical knowledge into machine learning models, and reviews major application areas, including chemistry acceleration, reduced order and manifold modeling, subgrid closure, and FDF closure. The author’s previous work on neural ordinary differential equation (NODE) based chemical kinetics and deep learning based FDF closure is discussed within the relevant subsections as representative applications of these physics-guided modeling principles. Section~\ref{sec:challenges} examines cross-cutting requirements for reliable deployment, including generalization, interpretability, uncertainty quantification, and solver integration, and identifies research directions motivated by these challenges. Section~\ref{sec:conclusion} concludes the article.

\section{Turbulent Combustion Modeling: Governing Equations and Conventional Closures}
\label{sec:closure}

\subsection{Governing equations and the closure problem}
\label{sec:governing}
Turbulent reacting flows are governed by the compressible reacting Navier-Stokes equations for mass, momentum, species, and energy. For a Newtonian mixture containing $N_s$ chemical species, the mass and momentum conservation equations are,
\begin{align}
\frac{\partial \rho}{\partial t} + \frac{\partial \rho u_j}{\partial x_j} &= 0,
\label{eq:continuity}\\[2pt]
\frac{\partial \rho u_i}{\partial t} + \frac{\partial \rho u_i u_j}{\partial x_j}
&= -\frac{\partial p}{\partial x_i} + \frac{\partial \sigma_{ij}}{\partial x_j},
\qquad
\sigma_{ij} = \mu\!\left(\frac{\partial u_i}{\partial x_j} + \frac{\partial u_j}{\partial x_i}
- \frac{2}{3}\,\delta_{ij}\frac{\partial u_k}{\partial x_k}\right),
\label{eq:momentum}
\end{align}
where $\mu$ is the dynamic viscosity. The species mass fractions evolve according to
\begin{equation}
\frac{\partial \rho Y_\alpha}{\partial t} + \frac{\partial \rho u_j Y_\alpha}{\partial x_j}
= \frac{\partial}{\partial x_j}\!\left(\rho D_\alpha \frac{\partial Y_\alpha}{\partial x_j}\right) + \rho\, \dot{\omega}_\alpha(\bm{\Phi}),
\qquad \alpha = 1,\dots,N_s,
\label{eq:species}
\end{equation}
where Fickian diffusion has been assumed for clarity, $\bm{\Phi}=(Y_1,\dots,Y_{N_s},h)$ denotes the vector of thermochemical scalars, and $\dot{\omega}_\alpha$ is the species chemical source term, typically expressed through Arrhenius kinetics. A representative energy equation, written for the specific enthalpy $h$ (sensible plus chemical) and neglecting radiation, is
\begin{equation}
\frac{\partial \rho h}{\partial t} + \frac{\partial \rho u_j h}{\partial x_j}
= \frac{\mathrm{D}p}{\mathrm{D}t}
+ \frac{\partial}{\partial x_j}\!\left(\lambda \frac{\partial T}{\partial x_j}
- \rho \sum_{\alpha=1}^{N_s} h_\alpha Y_\alpha V_{\alpha,j}\right)
+ \sigma_{ij}\frac{\partial u_i}{\partial x_j},
\label{eq:energy}
\end{equation}
where $\lambda$ is the thermal conductivity, $h_\alpha$ the species enthalpy, and $V_{\alpha,j}$ the species diffusion velocity. The system is closed by the ideal-gas equation of state. Equations~\eqref{eq:continuity}--\eqref{eq:energy} govern both the turbulence and the combustion, and both must be represented by any reduced description.

DNS resolves all scales of Eqs.~\eqref{eq:continuity}--\eqref{eq:energy}, but its cost grows prohibitively with Reynolds number, restricting it to canonical configurations \cite{chen2011}. Practical computations therefore solve averaged (RANS) or filtered (LES) equations, and LES has become the standard tool for high-fidelity combustor simulation because the unsteady large scale mixing that controls flame stabilization, ignition, and instabilities is resolved rather than modeled \cite{pitsch2006}. Applying a density weighted (Favre) filter, $\widetilde{\cdot} = \overline{\rho\,\cdot}/\bar{\rho}$, to the momentum equation~\eqref{eq:momentum} yields
\begin{equation}
\frac{\partial \bar{\rho}\widetilde{u}_i}{\partial t} + \frac{\partial \bar{\rho}\widetilde{u}_i\widetilde{u}_j}{\partial x_j}
= -\frac{\partial \bar{p}}{\partial x_i} + \frac{\partial \bar{\sigma}_{ij}}{\partial x_j}
- \frac{\partial \tau_{ij}^{\mathrm{sgs}}}{\partial x_j},
\qquad
\tau_{ij}^{\mathrm{sgs}} = \bar{\rho}\left(\widetilde{u_i u_j} - \widetilde{u}_i\widetilde{u}_j\right),
\label{eq:filteredmom}
\end{equation}
where $\tau_{ij}^{\mathrm{sgs}}$ is the SGS stress tensor; in RANS the analogous term is the Reynolds stress. These stress closures influence scalar mixing, flame stabilization, residence time, and wall heat transfer, so data-driven turbulence modeling is directly relevant to combustion even when the training data are non-reacting. Filtering the species equation~\eqref{eq:species} gives
\begin{equation}
\frac{\partial \bar{\rho}\widetilde{Y}_\alpha}{\partial t} + \frac{\partial \bar{\rho}\widetilde{u}_j\widetilde{Y}_\alpha}{\partial x_j}
= \frac{\partial}{\partial x_j}\!\left(\overline{\rho D_\alpha \frac{\partial Y_\alpha}{\partial x_j}}\right)
- \frac{\partial}{\partial x_j}\!\left[\bar{\rho}\left(\widetilde{u_j Y_\alpha} - \widetilde{u}_j\widetilde{Y}_\alpha\right)\right]
+ \bar{\rho}\,\widetilde{\dot{\omega}}_\alpha ,
\label{eq:filtered}
\end{equation}
in which three unclosed terms appear: the filtered molecular transport, the SGS scalar flux, and the filtered reaction rate $\widetilde{\dot{\omega}}_\alpha$. The flux terms can be closed using gradient diffusion models, but the source term is fundamentally harder. Because $\dot{\omega}_\alpha$ depends exponentially on temperature and nonlinearly on composition, the filtered rate cannot be evaluated at the filtered state:
\begin{equation}
\widetilde{\dot{\omega}}_\alpha(\bm{\Phi}) \neq \dot{\omega}_\alpha(\widetilde{\bm{\Phi}}),
\label{eq:tci}
\end{equation}
and evaluating the source term at filtered quantities can be in error by orders of magnitude \cite{poinsot2011}. Equation~\eqref{eq:tci} is a compact statement of the turbulence-chemistry interaction closure problem. Every turbulent combustion model is, at heart, a hypothesis for closing $\widetilde{\dot{\omega}}_\alpha$ and the associated mixing terms, and every ML model discussed in this article targets one or more of the unclosed terms in Eqs.~\eqref{eq:filteredmom} and \eqref{eq:filtered}. 

A general statistical resolution of Eq.~\eqref{eq:tci} is provided by the filtered density function, the one-point distribution of composition within a filter volume \cite{colucci1998,givi2006}. Given the FDF, $F(\bm{\psi};\bm{x},t)$, where  $\psi$ is the sample space variable corresponding to $\Phi$, any filtered function of composition $Q(\bm{\Phi})$ follows,
\begin{equation}
\widetilde{Q}(\bm{\Phi}) = \int Q(\bm{\psi})\, F(\bm{\psi};\bm{x},t)\, \mathrm{d}\bm{\psi},
\label{eq:fdfquad}
\end{equation}
so the chemical source term appears in closed form regardless of its nonlinearity. The transported FDF equation, however, contains its own unclosed conditional mixing term and is conventionally solved by Monte Carlo particle methods \cite{jaberi1999,haworth2010}. The associated cost motivates both the presumed-shape simplifications and the learned FDF closures developed in Section~\ref{sec:fdf}.

\subsection{Stiffness of detailed chemistry}
\label{sec:stiffness}
For a spatially homogeneous system, or within the reaction substep of an operator-split solver, the chemistry reduces to a coupled system of ordinary differential equations for the thermochemical state \cite{bansude2023},
\begin{equation}
\frac{\mathrm{d}\bm{\Phi}}{\mathrm{d}t} = \bm{S}(\bm{\Phi}),
\label{eq:kinetics}
\end{equation}
where the source terms $\mathbf{S}$ represent chemical reactions for the species equations and heat release for the enthalpy or temperature equation. The eigenvalue spectrum of Eq.~\eqref{eq:kinetics} spans many orders of magnitude: radical chain branching evolves on sub-microsecond scales, while fuel consumption and pollutant chemistry evolve far more slowly. Detailed mechanisms for real and alternative fuels involve hundreds to thousands of species and reactions \cite{lu2009,pope2013}, and this stiffness forces implicit, Jacobian-based integration whose factorization cost scales poorly with the number of species. Because Eq.~\eqref{eq:kinetics} must be advanced in every cell, or on every Monte Carlo particle, at every time step, chemistry integration commonly dominates the cost of reacting LES and DNS. Acceleration of this step, without loss of fidelity, is the second major target of data-driven methods, alongside closure of Eq.~\eqref{eq:tci}.

Any surrogate for Eq.~\eqref{eq:kinetics} must respect the constraints that the exact kinetics satisfy identically:
\begin{equation}
Y_\alpha \geq 0, \qquad \sum_{\alpha=1}^{N_s} Y_\alpha = 1, \qquad \mathbf{A}\,{\bm{\omega}} = \mathbf{0},
\label{eq:constraints}
\end{equation}
where $\mathbf{A}$ is the elemental composition matrix. Without these constraints, a network can produce small but persistent conservation errors that accumulate over millions of CFD time steps. This observation motivates much of the discussion in Section~\ref{sec:embedding}.

\subsection{Conventional closures}
\label{sec:conventional}


Conventional turbulent combustion models close Eq.~\eqref{eq:tci} by introducing structural assumptions about the unresolved interaction between turbulent mixing and chemical reaction. These models provide the physical foundation for many data-driven approaches: they identify the relevant variables, constraints, invariances, and closure quantities, and they define the baseline performance that learned models must improve upon or complement. Thus, PGML models are not developed independently of classical  combustion modeling; rather, it builds on these established closures by accelerating expensive components, correcting model-form errors, or learning specific unclosed terms from data. 

Mixing-controlled models, including the eddy dissipation concept (EDC) and the PaSR formulation, assume that reaction is fast relative to turbulent mixing, so the burning rate is proportional to a turbulent mixing frequency \cite{magnussen1977}. These models are robust and inexpensive, but they rely on empirical constants and degrade when finite-rate effects such as ignition, extinction, and emissions chemistry is present. Flamelet models instead view the turbulent flame as an ensemble of thin, locally one-dimensional laminar flame structures parameterized by a small set of control variables: mixture fraction $Z$ and scalar dissipation rate $\chi$ for non-premixed systems \cite{peters1984}, or a progress variable $c$ for premixed systems. Flamelet generated manifolds (FGM) \cite{vanoijen2000} and the flamelet/progress variable (FPV) formulation \cite{pierce2004} pre-tabulate the thermochemical state from canonical flame solutions; at run time only a few control variables are transported, and the table is interrogated with a presumed, typically $\beta$-shaped, subgrid PDF. Manifold methods substantially reduce computational cost, but their accuracy remains tied to the validity of the underlying manifold assumptions. In particular, the thin-flamelet approximation can break down for strongly unsteady, lifted, or highly diluted flames, while the size of the tabulated manifold increases rapidly as additional control variables are introduced. In premixed combustion, flame surface density (FSD) and thickened-flame models close subgrid wrinkling algebraically or via transport \cite{boger1998}.

Statistical closures represent TCI by evolving composition statistics rather than only mean scalar fields. Transported PDF methods solve a transport equation for the one-point joint PDF of the composition variables, and in some formulations also the velocity variables \cite{pope1985,haworth2010}. Their main advantage is that the chemical source term appears in closed form; therefore, no additional closure is required for the reaction term itself. However, molecular mixing remains unclosed and must be modeled. The LES counterpart is the filtered density function (FDF) methodology \cite{colucci1998,jaberi1999,givi2006}, including velocity--scalar FDF formulations \cite{sheikhi2007}. These methods are commonly solved using Lagrangian Monte Carlo particles or Eulerian stochastic fields. Their generality comes at a significant computational cost, since each notional particle carries a full thermochemical composition vector and may require repeated integration of stiff chemical kinetics.

Lower-cost alternatives include presumed-shape models, in which the sub-filter scalar distribution is specified analytically. The $\beta$ distribution is the standard choice for a conserved scalar because it is parameterized by the filtered mean and variance. However, this assumption can become inaccurate for multi-modal, skewed, narrow, or high-variance distributions. Conditional moment closure (CMC) provides another statistical approach by transporting conditional averages of reacting scalars conditioned on mixture fraction, based on the observation that fluctuations around the conditional mean are often smaller than unconditional fluctuations \cite{klimenko1999}. Capturing extinction, reignition, and strong conditional fluctuations, however, may require second-order or doubly conditioned CMC formulations, increasing model complexity and computational cost.

In parallel with these closure strategies, the chemical mechanism itself can be simplified to reduce stiffness and dimensionality. Intrinsic low-dimensional manifolds exploit time-scale separation in composition space \cite{maas1992}, in situ adaptive tabulation (ISAT) stores chemistry information on demand in regions visited by the solver \cite{pope1997}, and graph-based skeletal reduction removes less important species and reactions from detailed mechanisms \cite{lu2009}. These classical reduction and tabulation methods provide important physical and algorithmic foundations for the learned manifold and chemistry-acceleration approaches discussed in Section~\ref{sec:rom}.


Across conventional turbulent combustion closures, three difficulties recur. The first is the high cost of stiff chemical kinetics, which is especially significant in finite-rate chemistry, PDF/FDF, and CMC calculations. The second is the reliance on restrictive closure assumptions, including presumed PDF shapes, modeled micro-mixing rates, and neglect of conditional fluctuations; the associated empirical constants often do not transfer reliably across combustion regimes. The third is the rapid increase in the dimensionality and storage cost of tabulated manifolds as additional control variables are introduced. These challenges align closely with the strengths of modern machine learning: approximating expensive nonlinear mappings, identifying low-dimensional structure in high-dimensional data, and replacing empirical closure functions with flexible learned representations \cite{ihme2022}. The next section discusses how physics-guided learning can use the physical structure summarized here to address both the TCI  closure problem in Eq.~\eqref{eq:tci} and the cost of chemical integration in Eq.~\eqref{eq:kinetics}.

\section{Physics-Guided Machine Learning for Turbulent Combustion}
\label{sec:pgml}


This section presents data-driven and physics-guided modeling for turbulent combustion in a unified framework. Section~\ref{sec:formulation} first defines the learning problem and discusses the distinction between a priori and a posteriori assessment. Section~\ref{sec:embedding} then describes how physical knowledge can be incorporated into machine learning models through data design, input features, model architecture, loss functions, hybrid formulations, and solver-aware validation. Sections~\ref{sec:chemacc}-\ref{sec:fdf} discuss the major application areas: chemistry acceleration, reduced order and manifold modeling, subgrid closure, and FDF closure. Within these discussions, the author's work on neural ordinary differential equations-based chemical kinetics and deep-learning-based FDF closure is presented as representative examples of physics-guided modeling for combustion.

\subsection{Formulating the learning problem and its assessment}
\label{sec:formulation}


Most data-driven combustion closure problems can be formulated as supervised regression tasks. Given a set of input features $\bm{x}$, such as filtered moments, control variables, or local thermochemical states, and target quantities $\bm{y}$, such as exact unclosed terms extracted from high-fidelity data or time advanced states, the model parameters are obtained by solving
\begin{equation}
\bm{\theta}^* = \arg\min_{\bm{\theta}} \sum_n \mathcal{L}\!\left(f_{\bm{\theta}}(\bm{x}_n), \bm{y}_n\right),
\label{eq:supervised}
\end{equation}
where $\mathcal{L}$ is commonly a mean-squared or mean-absolute error, often supplemented with regularization to improve generalization and reduce overfitting.

Different machine learning architectures are suited to different combustion modeling tasks. Multilayer perceptrons (MLP) are commonly used for pointwise thermochemical mappings, such as source term or table-interpolation models. Convolutional neural networks (CNNs) exploit spatial locality and translation equivariance, making them suitable for field-to-field closure modeling on structured LES data. Neural ordinary differential equations (ODEs) \cite{chen2018} represent the network as the right-hand side of a differential equation, which is particularly natural for dynamical systems such as chemical kinetics. Operator-learning methods, including DeepONet \cite{lu2021} and Fourier neural operators \cite{li2021}, learn mappings between functions and are therefore useful for parametric field prediction and surrogate modeling. Physics-informed neural networks (PINNs)\cite{raissi2019} incorporate governing equation residuals directly into the training objective. In unsupervised or reduced order settings, principal component analysis and autoencoders can identify low-dimensional composition manifolds from high-dimensional thermochemical data \cite{sutherland2009}. Broader discussions of these approaches in combustion, fluid mechanics, and scientific machine learning are available in \cite{ihme2022,zhou2022,brunton2020,duraisamy2019}.

Three practical issues are central to the development of data-driven combustion models. The first is data generation: for subgrid closure modeling, training targets are commonly obtained by explicitly filtering DNS fields at one or more filter widths, while chemistry surrogates typically use data from canonical reactors, laminar flames, flamelet libraries, or sampled thermochemical manifolds. The sampling strategy determines which regions of composition space are represented in the training set and therefore defines the effective domain of validity of the learned model. Since reacting states occupy a low-dimensional region of the full composition space, the distribution and coverage of the training data are often more important than the total number of samples, and multi-scale sampling has been shown to improve accuracy across the widely varying magnitudes of chemical kinetic rates \cite{zhang2022}. The second issue is feature selection: model inputs should correspond to quantities available to the CFD solver at run time, such as filtered scalars, scalar variances, local thermochemical variables, turbulence time scales, or control variables used in manifold models. These inputs should be scaled or non-dimensionalized and selected with attention to relevant invariances, as discussed in Section~\ref{sec:embedding}. Non-dimensionalization using physically meaningful scales, such as filter width, laminar flame speed, flame thickness, and chemical time scales, improves numerical conditioning and encodes similarity across operating conditions; for example, models expressed in terms of Karlovitz- and Damk\"ohler-number-like groups are more likely to transfer across related conditions than models trained only on dimensional variables. The third issue is validation: random splitting of spatially or temporally correlated data can leak information between training and test sets and lead to overly optimistic error estimates. More reliable assessment requires testing on withheld physical conditions rather than merely withheld data points, such as held-out flames, filter widths, fuels, pressures, equivalence ratios, or operating conditions. Such validation protocols provide a more meaningful measure of model generalization and are essential before a learned closure or chemistry surrogate is used in coupled CFD simulations \cite{ihme2022}.

\begin{figure}[t]
\centering


\includegraphics[width=0.8\textwidth]{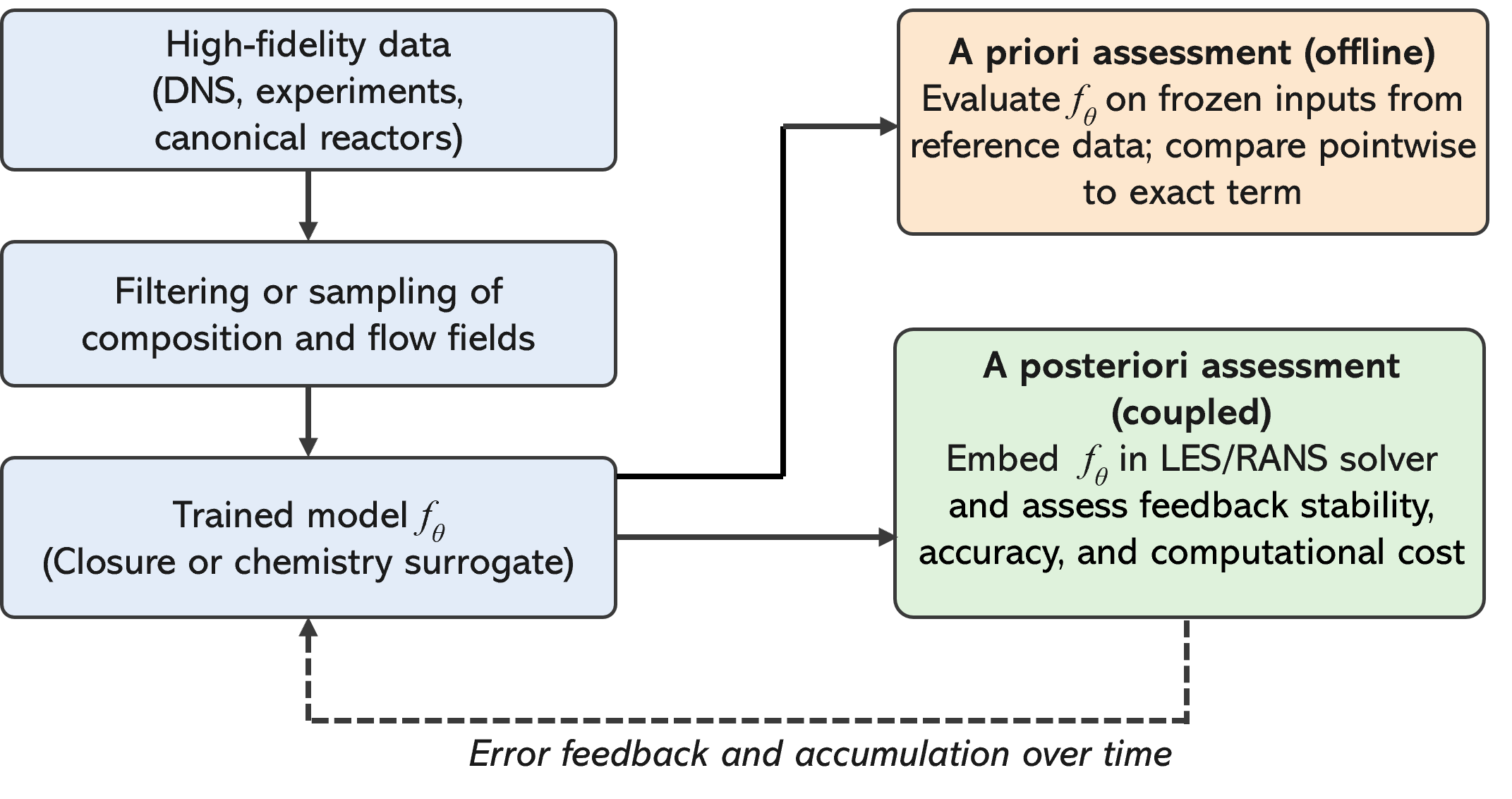}

\caption{A priori (offline) versus a posteriori (coupled) assessment of data-driven combustion models. In a priori testing the model sees frozen inputs from reference data; in a posteriori testing its errors feed back through the solver, so stability and error accumulation, not only pointwise accuracy, determine success.}
\label{fig:apriori}
\end{figure}

Equally fundamental is the distinction between a priori and a posteriori assessment, illustrated in Fig.~\ref{fig:apriori}. In a priori testing the trained model is evaluated offline on frozen inputs taken from the reference data and compared pointwise with the exact unclosed term. In a posteriori testing the model is coupled to the flow solver, so its errors feed back into the resolved fields and can accumulate or destabilize the computation. Experience in both non-reacting and reacting LES shows that good a priori correlation does not guarantee a posteriori robustness. A network trained to minimize pointwise SGS error may inject energy into unresolved scales or provide insufficient dissipation, and the resulting LES can become unstable even when the offline error is low \cite{beck2019,maulik2019,duraisamy2019}. In combustion, instability may also arise through species overshoots, negative temperatures, or spurious heat release. Data-driven closures that outperform classical models offline can therefore require regularization, projection onto stable model forms, or constrained outputs to run stably when coupled. A posteriori demonstration in a configuration not used for training is consequently regarded as the meaningful test of a combustion model, and coupled diagnostics such as kinetic-energy and scalar-variance budgets, boundedness, conservation over long time horizons, and solver robustness, belong alongside pointwise regression error in any evaluation.

\subsection{Embedding physical knowledge}
\label{sec:embedding}

The failure modes of unguided regression follow directly from the structure of the problem \cite{ihme2022,duraisamy2019,beck2019}. A network predicting species production rates independently will not, in general, satisfy the element balance of Eq.~\eqref{eq:constraints}, and the resulting mass and element drift accumulates over the course of a simulation. Predicted mass fractions may leave $[0,1]$, variances may become negative, and predicted Reynolds or SGS stresses may be non-realizable, producing states from which a compressible solver rarely recovers. Independently predicted temperature and composition may be incompatible with the enthalpy-temperature relation or the equation of state, corrupting the pressure field. Because training data occupy a thin manifold in composition space, a trajectory that drifts off-manifold receives essentially arbitrary predictions whose errors compound autoregressively. Finally, as discussed in Section~\ref{sec:formulation}, a closure that is accurate against filtered-DNS targets can still destabilize an LES because it was never exposed to the numerics, resolution, and feedback of the deployed solver.

\begin{figure}[t]
\centering

\includegraphics[width=\textwidth]{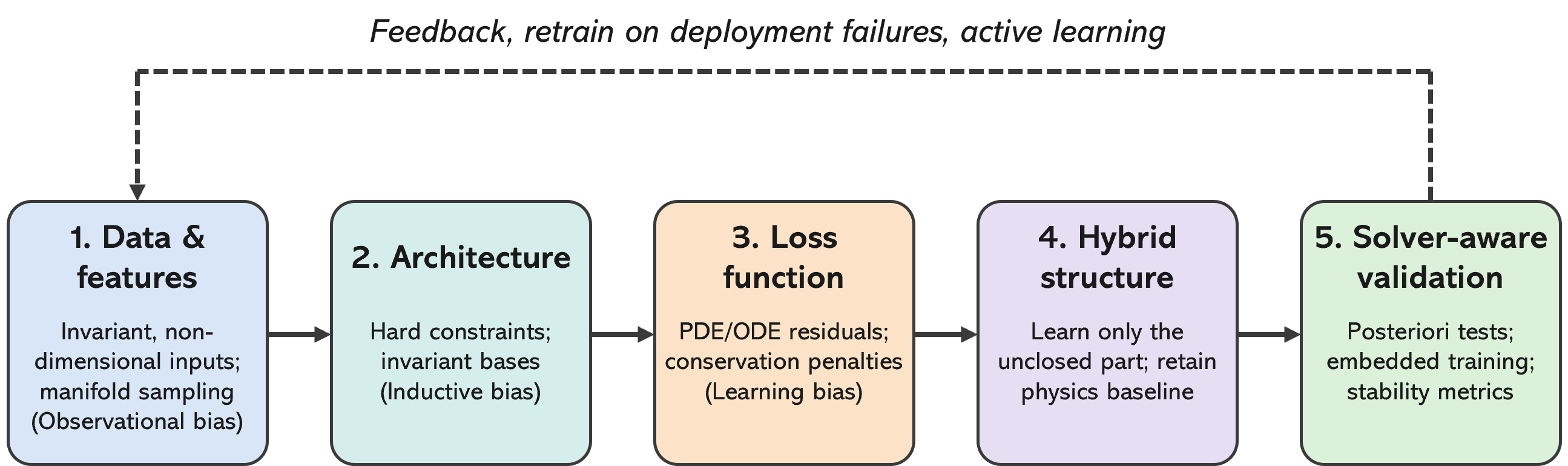}
\caption{Mechanisms for incorporating physical knowledge into a PGML pipeline for combustion modeling, following the taxonomy of Karniadakis et al.~\cite{karniadakis2021}. Physical constraints and domain knowledge can be introduced through data generation and feature selection, model architecture, loss-function design, hybrid closure formulation, and solver-aware validation. The feedback loop indicates retraining informed by a posteriori solver behavior.}
\label{fig:routes}
\end{figure}

Physics-guided machine learning addresses these failure modes by incorporating prior knowledge at five stages of the modeling pipeline, as illustrated in Fig.~\ref{fig:routes}: data generation and feature selection, model architecture, loss-function design, hybrid model formulation, and solver-aware training and validation. This organization is consistent with the taxonomy of Karniadakis et al.~\cite{karniadakis2021} and its combustion focused discussion by Ihme et al.~\cite{ihme2022}. In this taxonomy, \emph{observational bias} refers to the use of physics in data design and feature selection, \emph{inductive bias} refers to physical structure embedded in the model architecture, and \emph{learning bias} refers to physics-based terms introduced through the training objective. Hybrid modeling and solver-aware validation provide two additional mechanisms that are especially important for CFD applications, where learned closures must interact robustly with discretization, time integration, and feedback from the flow solver. In practice, reliable combustion ML models often combine several of these mechanisms. The five stages are discussed below.

\paragraph{Data and input features.} One of the most effective ways to introduce physical guidance is through the construction of the training data and the selection of input features. The inputs should respect the symmetries and invariances of the modeled quantity. For example, in turbulence modeling, Galilean invariance favors velocity gradients over raw velocity components, while rotational invariance motivates the use of tensor invariants rather than individual tensor components \cite{duraisamy2019,ling2016}. Non-dimensionalization also plays an important role because it encodes similarity across operating conditions and improves numerical conditioning, as discussed in Section~\ref{sec:formulation}. Equally important is the sampling of the thermochemical state space. The training data should cover the composition states that the solver is likely to encounter, including off-manifold states generated by molecular mixing. For this reason, chemistry-tabulation studies have used composite sampling strategies, including canonical flames with randomized perturbations \cite{ding2021}, stochastic micro-mixing emulation \cite{wan2020}, constrained-equilibrium sampling \cite{chatzopoulos2013}, and multi-scale sampling across reaction-rate magnitudes \cite{zhang2022}. For closure modeling based on filtered DNS, the training targets should be generated using filter and grid operations that are consistent with the deployed LES, reducing mismatch between a priori assessment and a posteriori performance \cite{beck2019}. When appropriate, symmetry-preserving data augmentation, such as rotations or reflections of DNS fields, can further improve robustness without violating the physical structure of the target problem.

\paragraph{Architecture.} Physical constraints can also be hard coded directly in the model architecture. This introduces an inductive bias, meaning that the model is designed to satisfy selected physical properties before training begins. Architectural constraints are particularly valuable because they are enforced by construction and therefore remain active during deployment, unlike soft penalties that are satisfied only approximately \cite{karniadakis2021}. In turbulence modeling, for example, tensor basis neural network (TBNN) express the predicted anisotropy tensor using invariant tensor bases constructed from the resolved strain-rate and rotation-rate tensors. This guarantees Galilean and rotational invariance and has influenced related SGS stress and scalar-flux closures for reacting LES \cite{ling2016,chung2022}. Similar architectural ideas are useful in combustion chemistry. Bounded transformations can restrict predicted mass fractions or progress variables to physically admissible ranges. Normalized output layers can ensure that species mass fractions remain non-negative and sum to unity. Element conservation can be enforced by predicting only element-conserving increments, projecting source terms onto the null space of the elemental composition matrix, or reconstructing species production rates from reaction progress rates through the stoichiometric matrix. The chemical reaction neural network (CRNN) follows this idea by arranging network parameters so that they correspond to stoichiometric coefficients, reaction orders, and rate constants, making the learned model interpretable and conservative by design \cite{ji2021crnn}. For chemical kinetics, another useful architectural choice is to represent the learned model as the right-hand side of an ordinary differential equation. This neural ODE formulation preserves the continuous-time dynamical-system structure of chemistry evolution, avoids dependence on a fixed time step, and allows the use of standard time integration and sensitivity-analysis tools \cite{chen2018,owoyele2022,kumar2025}.

\paragraph{Loss function.} When physical constraints cannot be enforced directly through the architecture, they can be incorporated into the training objective as soft penalties:
\begin{equation}
\mathcal{L} = \mathcal{L}_{\mathrm{data}}
+ \sum_k \lambda_k\, \mathcal{L}_{\mathrm{phys},k},
\label{eq:softloss}
\end{equation}
where $\mathcal{L}_{\mathrm{data}}$ measures agreement with the training data and each $\mathcal{L}_{\mathrm{phys},k}$ penalizes violation of a physical constraint. These penalties may correspond to governing equation residuals, conservation errors, boundedness or realizability violations, or inconsistencies between predicted and transported moments. The weights $\lambda_k$ control the relative importance of the different physical penalties.  Physics-informed neural networks provide a common example of this strategy. In PINNs, automatic differentiation is used to evaluate governing equation residuals at collocation points, allowing the model to be constrained by the differential equations even in regions where data are sparse or unavailable \cite{raissi2019,karniadakis2021}. This approach is flexible because it does not require a specialized model architecture, but the constraints are satisfied only approximately. The penalty weights must be chosen carefully, and the resulting optimization problem can become ill-conditioned when different loss terms have widely different magnitudes or gradients. Such gradient imbalance and training difficulties are well-documented in stiff and multi-scale problems \cite{wang2021,krishnapriyan2021}. These issues are particularly important in combustion, where target quantities may vary by many orders of magnitude. For example, an absolute error loss can underweight trace radicals that control ignition, while a purely relative error can overemphasize numerical noise in species with nearly zero concentration. Loss functions should therefore be designed around the quantities that determine physical accuracy, such as ignition delay, flame speed, extinction limits, heat release \textit{etc}.. In stiff chemical kinetics, direct PINN formulations can struggle because of the widely separated chemical time scales. This motivated approaches such as Stiff-PINN, where quasi-steady-state assumptions from classical kinetics reduction are used to reduce the stiffness before applying the PINN formulation \cite{ji2021}.  In general, hard constraints are preferable for properties whose violation can destabilize a coupled CFD simulation, such as boundedness, normalization, element conservation, and invariance. Soft penalties are most useful for regularization, inverse problems, weakly enforced physical preferences, and constraints that are difficult to impose directly through the architecture.

\paragraph{Hybrid structure.} Hybrid modeling strategies retain the structure of a physics-based baseline model while using machine learning to represent only the missing closure term or model-form correction. This approach is often more robust than replacing the full model with a black-box surrogate, because the governing conservation laws, boundary conditions, transport equations, and numerical coupling remain handled by the CFD solver. The learned component is restricted to the part of the model that is uncertain or empirically closed, which reduces the hypothesis space and improves interpretability. Examples include field inversion and machine learning, where a spatially varying discrepancy in a RANS transport equation is first inferred and then learned as a function of local flow features \cite{parish2016}; Reynolds stress discrepancy modeling in physics-constrained coordinates \cite{wang2017}; flamelet-based models in which the presumed sub-filter PDF \cite{henrydefrahan2019} or scalar dissipation rate input \cite{yellapantula2021} is replaced by a learned model while the manifold formulation is retained; and CNN-based closures that provide filtered reaction rates within otherwise conventional premixed LES formulations \cite{lapeyre2019,seltz2019}. In these examples, the learned component is localized, physically anchored, and dimensionally consistent, which improves the likelihood of stable a posteriori behavior \cite{duraisamy2019}. The learned FDF closure discussed in Section~\ref{sec:fdf} follows this hybrid modeling philosophy: the resolved transport equations are retained, while the sub-filter distribution is supplied by a data-driven model.

\paragraph{Solver-aware training and validation.} The final mechanism is the deployment environment itself. In CFD, the learned model does not operate in isolation; it interacts with the discretization, filtering operation, time-integration scheme, density feedback, and numerical stabilization used by the solver. These factors can strongly influence whether a model that performs well in offline testing remains accurate and stable in coupled simulations. Three practices are particularly important. First, a posteriori testing should be treated as a primary validation step. Learned closures must be evaluated inside running LES or RANS calculations, where their errors can feed back into the resolved fields and affect long-time statistics, conservation, and numerical stability \cite{franke2017,readshaw2021,lapeyre2019}. Second, embedded or differentiable-physics training can be used when the solver is made differentiable and the loss is defined on the evolved solution rather than on the instantaneous closure target. This allows gradients to propagate through multiple solver steps and has been shown to improve deployment stability in data-driven closure modeling \cite{sirignano2020,macart2021}. Third, stability-oriented training is important for autoregressive or time-evolving chemistry models. Such approaches train on trajectories rather than isolated time steps, introduce perturbations to mimic deployment drift, and encourage stable behavior near equilibrium \cite{owoyele2022,kim2021}. This is especially relevant for learned chemical-kinetics models, including neural ODE-based kinetics surrogates, because small trajectory errors can accumulate during repeated chemistry integration; this class of models is discussed in detail in Section~\ref{sec:chemacc}. The feedback loop in Fig.~\ref{fig:routes} can therefore be interpreted as an active-learning process: the model is monitored during deployment, failure states are identified, and additional data are generated to improve the training set. This philosophy is closely related to the on-demand nature of ISAT, where chemistry information is accumulated in regions of composition space visited by the solver \cite{pope1997}. A staged validation hierarchy is also essential. Chemistry surrogates should first be tested in homogeneous reactors, then in one-dimensional flames or mixing reactors, and only then in turbulent reacting flow simulations. Similarly, an FDF closure should first reproduce conserved-scalar mixing before being used in finite-rate reacting LES. This hierarchy helps isolate model errors before they are amplified in fully coupled CFD calculations and is followed in the case studies discussed below.

\subsection{Combustion chemistry representation and acceleration}
\label{sec:chemacc}

Combustion chemistry representation and acceleration is one of the most mature applications of machine learning in combustion. The objective is to replace, accelerate, or approximate the expensive integration of chemical kinetics or the lookup of tabulated thermochemical states. Given a thermochemical composition vector $\bm{\Phi}(t)$, the learned target may be the reaction-rate vector ${\bm{\omega}}(\bm{\Phi})$ or, in time-advancement formulations, the updated state after an operator-splitting chemistry step. Early work in this direction predates modern deep learning: Christo et al.~\cite{christo1996} used artificial neural networks to represent reduced H$_2$/CO$_2$ chemistry within a transported-PDF computation, and Ihme et al.~\cite{ihme2009} later showed that optimized neural-network tabulation could outperform structured tabulation in LES of a bluff-body swirl-stabilized flame. A central difficulty in these models is coverage of composition space. A network trained only on states from a limited flame family may become inaccurate when it encounters thermochemical states outside that training distribution, often without providing a clear indication of failure. Several studies have addressed this issue through improved sampling strategies. Chatzopoulos, Franke, and co-workers trained neural networks on abstract thermochemical problems designed to span the accessible composition region, including rate-controlled constrained-equilibrium states and flamelet ensembles, and deployed them in LES--PDF simulations of turbulent non-premixed flames, including Sydney flame L with strong local extinction \cite{chatzopoulos2013,franke2017}. Ding et al.~\cite{ding2021} and Readshaw et al.~\cite{readshaw2021} further introduced hybrid flamelet/random-data sampling with multiple multilayer perceptrons, obtaining errors comparable to direct integration at a reduced cost in RANS--PDF and stochastic-fields LES--PDF simulations of the Sandia flames D--F. Wan et al.~\cite{wan2020} used stochastic micro-mixing trajectories to generate training data representative of the mixing environment in the target flame and applied the resulting model in DNS of a syngas oxy-flame with wall effects. Multi-scale sampling has also been shown to improve accuracy across the widely varying magnitudes of chemical kinetic rates \cite{zhang2022}. A closely related direction focuses on compressing flamelet or manifold tables rather than replacing the kinetic source terms directly. In this setting, neural networks serve as smooth, differentiable, and memory-efficient approximations of physics-based thermochemical manifolds, reducing the cost and storage requirements of multidimensional table lookup while retaining the underlying flamelet or manifold formulation \cite{ihme2009,zhang2020fgm}.

Continuous-time formulations provide another approach for learning chemical kinetics. Instead of predicting a discrete state update over a prescribed time step, these methods learn the right-hand side of the chemical evolution equation directly. In the neural ODE formulation \cite{chen2018}, the chemical source term is represented as
\begin{equation}
\frac{\mathrm{d}\bm{\Phi}}{\mathrm{d}t} = f_{\bm{\theta}}(\bm{\Phi}),
\label{eq:node}
\end{equation}
where $f_{\bm{\theta}}$ is a neural-network approximation of the kinetic source term vector. The model is trained so that the trajectories obtained by integrating Eq.~\eqref{eq:node} match reference trajectories generated from the detailed chemical mechanism. Gradients can be computed either by backpropagation through the time integrator or by adjoint sensitivity analysis. This inductive bias preserves the continuous-time dynamical structure and decouples the model from any particular time step.  The principal obstacle is stiffness. Chemical systems contain fast radical reactions and slower fuel-consumption or pollutant-formation pathways, leading to widely separated time scales. As a result, explicit integration during training may require very small time steps, and gradients can become ill-conditioned. Kim et al.~\cite{kim2021} examined these issues for stiff neural ODEs and showed that implicit integration, adjoint sensitivity analysis, and appropriate scaling of species and time can make training more tractable. The scaling is physically meaningful because major species, minor species, and radicals occur at very different concentration levels. Several related approaches have extended this idea. ChemNODE demonstrated learned source term models for hydrogen and small-hydrocarbon autoignition over ranges of temperature, and equivalence ratio \cite{owoyele2022}. Physics-constrained neural ODE formulations further incorporate conservation penalties to improve robustness \cite{kumar2025}. The CRNN provides a complementary strategy by embedding the structure of a reaction mechanism directly into the network architecture \cite{ji2021crnn}. Operator-learning methods, including extended deep operator networks \cite{goswami2024} and Fourier neural operators \cite{weng2025}, extend chemistry acceleration across families of initial conditions and operating parameters, offering an alternative to repeated time integration during deployment.

The author's work on neural ODE-based chemical kinetics \cite{bansude2022,bansude2023} follows this continuous-time modeling strategy and illustrates how physics-guided design choices can improve both training stability and deployment performance. The first study developed a data-driven NODE chemistry integrator for a constant-pressure homogeneous hydrogen-air reactor described using $9$ species, $21$ step finite-rate mechanism. The problem was formulated in an operator-splitting context, where the chemical reaction step is separated from transport or mixing and treated as a local system of stiff ODEs. Instead of learning a single-step state update, the NODE was trained to represent the chemical source term vector and was integrated along the thermochemical trajectory. This formulation directly targets multi-step accuracy, which is essential because chemistry surrogates are used repeatedly during CFD calculations. Two physics-guided elements were central to the framework. First, systematic sampling of the composition space was used to expose the network to thermochemical states encountered over the relevant ranges of initial temperature and equivalence ratio. Second, normalization of species mass fractions and temperature regularized the large differences in variable magnitudes, including the disparity between major species and trace radicals, and thereby stabilized training of the stiff chemical system. By embedding an explicit ODE solver in the NODE training loop, the learned source term representation became less stiff than the original detailed mechanism, allowing accurate integration with explicit solvers during deployment. The resulting model captured ignition trajectories, species evolution, and ignition-delay trends over the tested conditions while reducing multi-step error accumulation relative to conventional single-step regression-type chemistry surrogates.

\begin{figure}[t]
\centering
\includegraphics[width=\textwidth]{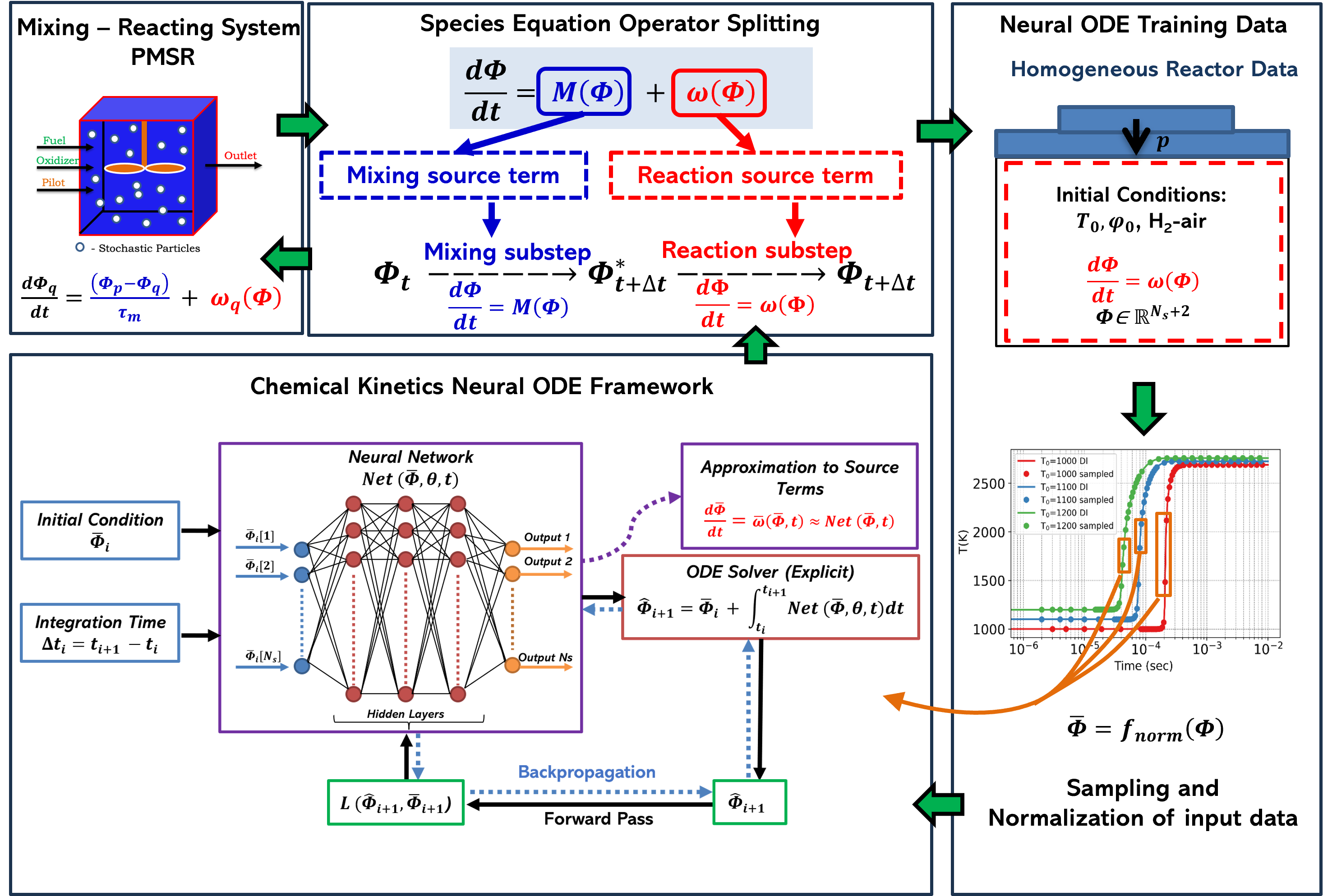}
\caption{Physics-guided neural ODE framework for chemical-kinetics acceleration in reacting flow simulations. The reacting system is formulated through operator splitting, in which the thermochemical state evolves through separate mixing and reaction substeps. The reaction substep is represented as a continuous-time neural ODE, trained using systematically sampled and normalized homogeneous-reactor trajectories. The embedded explicit ODE solver, trajectory-level loss, and backpropagation through the integrated solution enable the network to learn a reduced-stiffness representation of the chemical source term. The trained NODE surrogate can then be coupled with mixing models, such as the pairwise mixing stirred reactor, to assess stability, multi-step accuracy, and computational efficiency under mixing-induced perturbations \cite{bansude2022,bansude2023}.}
\label{fig:nodeworkflow}
\end{figure}

\begin{figure}[t]
\centering
\includegraphics[width=0.98\textwidth]{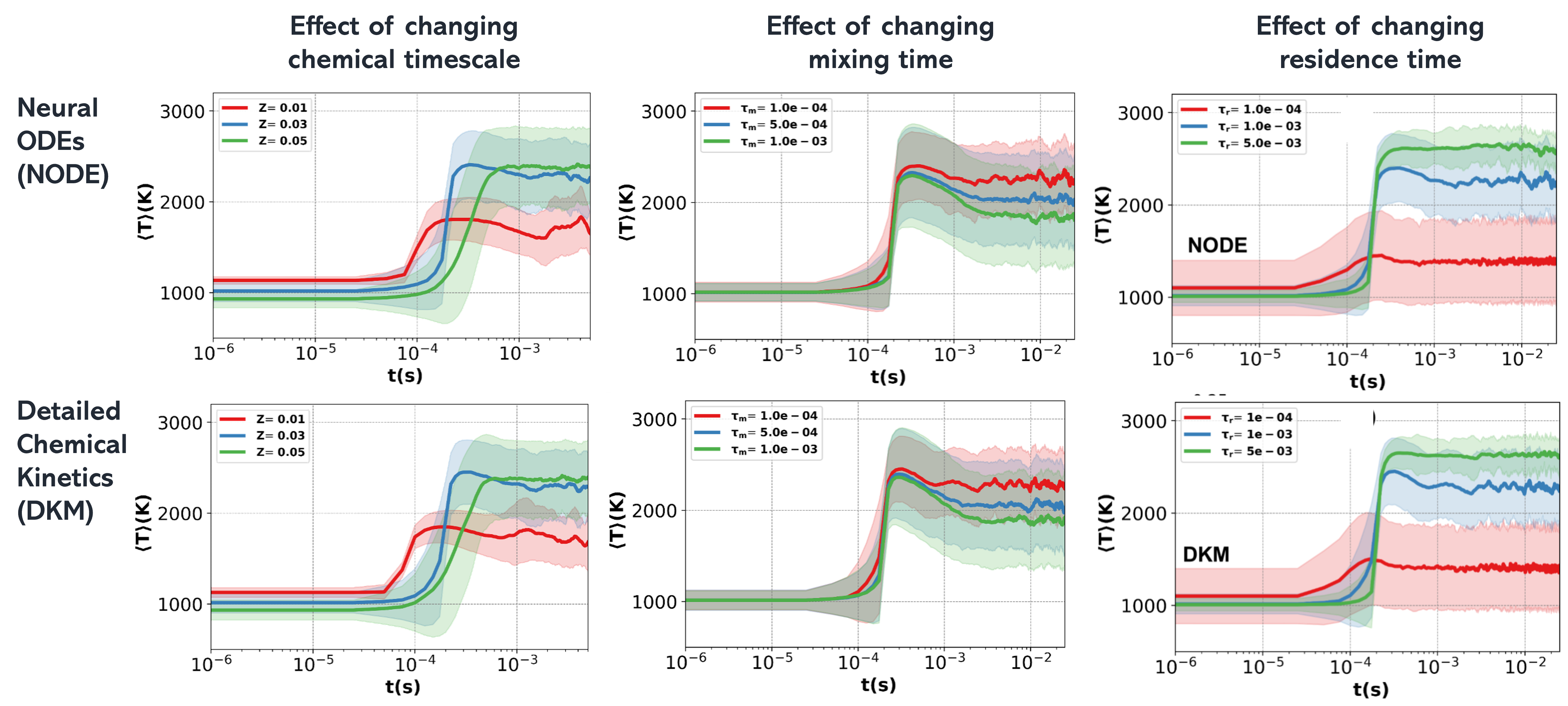}
\caption{PMSR assessment of the neural ODE (NODE) chemical-kinetics surrogate. The plots compare the stochastic-particle ensemble averaged temperature, $\langle T\rangle$, obtained using the NODE surrogate (top row) and direct integration of the detailed kinetic mechanism (DKM; bottom row). The shaded regions denote RMS temperature fluctuations across the particle ensemble, reflecting the variability introduced by stochastic mixing. Results are shown for variations in chemical time scale, mixing time, and residence time. The close agreement between NODE and DKM in both the mean temperature evolution and fluctuation levels demonstrates the ability of the NODE surrogate to generalize across coupled mixing-reaction conditions \cite{bansude2023}.}
\label{fig:pmsrresults}
\end{figure}

The second study extended this assessment from the homogeneous reactor to a pairwise mixing stirred reactor (PMSR), which provides a zero-dimensional stochastic mixing-reaction configuration relevant to transported PDF and FDF combustion solvers \cite{bansude2023}. This step is important because a chemistry surrogate that performs well in an isolated reactor may still fail when coupled with mixing, where particles or notional fluid elements are repeatedly driven away from the reaction manifold. In the PMSR formulation, mixing and reaction were advanced through an operator-splitting procedure, as illustrated in Fig.~\ref{fig:nodeworkflow}, allowing the NODE chemistry model to be tested under repeated mixing-induced perturbations and across different mixing and chemical time scales. The NODE retained stable and accurate predictions of global thermochemical evolution in the presence of mixing, demonstrating improved robustness beyond the training configuration of homogeneous-reactor trajectories, as shown in Fig.~\ref{fig:pmsrresults}. The PMSR assessment also showed that the trained NODE was not restricted to the ODE solver used during training. Because the NODE represents the continuous source term rather than a fixed time-step map, it could be integrated with different solvers and tolerances during deployment. This solver flexibility produced larger computational gains than in the homogeneous-reactor study: by using suitable explicit solvers and relaxed tolerances while maintaining comparable accuracy in global combustion quantities, the PMSR calculations achieved up to an order-of-magnitude speed-up relative to direct integration of the detailed kinetic mechanism. Together, these two studies show how observational bias through systematic data generation, normalization-based regularization, ODE-based architectural bias, trajectory-level training, and solver-aware validation can be combined to construct a physics-guided chemistry surrogate suitable for eventual coupling with turbulent combustion simulations.


\subsection{Reduced-order modeling and manifold learning}
\label{sec:rom}

Chemistry acceleration and manifold modeling address the same computational bottleneck from complementary perspectives. Instead of learning the source terms for all $N_s$ species in the full composition space, reduced order approaches seek a lower-dimensional representation of the thermochemical state. Classical methods identify this reduced representation using time-scale analysis of the chemical mechanism, as in intrinsic low-dimensional manifolds \cite{maas1992}. Data-driven approaches instead infer the manifold directly from composition data. Sutherland and Parente introduced PCA-based combustion modeling, in which a small number of principal components are transported and their chemical source terms are learned from data \cite{sutherland2009}. Subsequent work coupled PCA with Gaussian-process regression to perform LES principal-component-transport simulations of the Sandia flames D, E, and F, including cases with extinction and reignition at higher jet velocities \cite{malik2021}. Ranade and Echekki further developed a data-based turbulent combustion closure framework in which both chemical source terms and unresolved mixing contributions are represented in the reduced space \cite{ranade2019}. Autoencoders extend this idea beyond linear PCA by learning nonlinear latent representations of thermochemical composition.

Learned manifolds play a role similar to flamelet-based manifolds, but the reduced coordinates are inferred from data rather than prescribed from canonical flame solutions. This provides additional flexibility, but it can also reduce interpretability. For a reduced representation to be useful in CFD, the latent variables must remain compatible with transport equations, boundary conditions, and physical constraints such as mass and element conservation. A practical strategy is therefore to combine physics-defined coordinates with learned corrections. For example, mixture fraction and progress variable may be retained as primary coordinates, while an autoencoder captures additional degrees of freedom associated with heat loss, differential diffusion, pressure variation, or finite-rate effects. This type of hybrid manifold construction is particularly relevant for combustion applications in which differential diffusion, ignition sensitivity, heat-loss effects, or pollutant chemistry can limit the accuracy of simple low-dimensional manifolds. Evolving the latent-space dynamics using neural ODEs provides a natural connection between reduced order manifold learning and chemistry acceleration, and represents an important direction for physics-guided combustion modeling.

\subsection{Subgrid scale closure modeling and augmentation}
\label{sec:sgs}

A related area of development addresses the unclosed terms of Eqs.~\eqref{eq:filteredmom} and \eqref{eq:filtered} directly, almost always through supervised training on filtered DNS: a DNS field is filtered to LES resolution, the exact unclosed term is computed, and the term is regressed from resolved-scale inputs. Developments in non-reacting turbulence modeling directly preceded and informed the combustion modeling. For example, tensor basis neural networks introduced invariance-preserving architectures for Reynolds-stress modeling \cite{ling2016}, while field inversion and machine learning \cite{parish2016} and Reynolds-stress discrepancy reconstruction \cite{wang2017} established the idea of learning model-form corrections rather than replacing the full turbulence model. Other studies introduced prediction-confidence measures based on distance from the training manifold \cite{wu2017}, Bayesian approaches for model-form uncertainty \cite{geneva2019}, and sparse symbolic regression for interpretable algebraic closures \cite{schmelzer2020}. These developments are directly relevant to reacting flow LES because SGS stresses, scalar fluxes, and scalar dissipation control mixing, residence time, flame stabilization, and TCI. For LES closure modeling, an important challenge is the mismatch between the filtered DNS data used for training and the numerical environment in which the closure is deployed. Beck et al.~\cite{beck2019} examined this issue for data-driven LES closures, while Maulik et al.~\cite{maulik2019} showed that good a priori agreement with filtered-DNS targets does not necessarily lead to stable a posteriori simulations. Sirignano, MacArt, and Freund addressed this limitation by training closures inside the discretized solver using differentiable programming, so that the learned model is optimized based on its effect on the evolved solution rather than only its instantaneous closure error \cite{sirignano2020,macart2021}. These studies highlight three principles that are especially important for combustion closure modeling: the input representation should respect invariance and realizability constraints, the model should include a measure of prediction confidence or out-of-distribution behavior, and the final assessment must be based on performance when the closure is propagated through the governing equations.

These principles have also shaped machine learning closures for reacting flow LES. In premixed combustion, CNN is particularly useful because subgrid flame wrinkling is inherently spatial and multi-scale. Quantities such as flame curvature, orientation, and unresolved flame surface area cannot be inferred reliably from a single local value of the progress variable. Lapeyre et al.~\cite{lapeyre2019} trained a CNN using filtered DNS of slot-burner flames to predict the filtered flame surface density and, consequently, the SGS reaction rate. Their model outperformed algebraic and dynamic wrinkling closures in a priori assessment. Seltz et al.~\cite{seltz2019} used a related strategy to map resolved LES quantities directly to the terms in a filtered flamelet generated manifold balance equation, thereby closing both the progress variable source term and unresolved transport. Nikolaou et al.~\cite{nikolaou2019} combined CNN-based deconvolution with explicit filtering to reconstruct the unfiltered progress variable, from which the subgrid variance and filtered reaction rate were obtained. Other studies have focused on specific closure quantities required by flamelet and mixing models. For example, Yellapantula et al.~\cite{yellapantula2021} learned the filtered progress variable dissipation rate from filtered DNS of premixed hydrogen flames over a range of Karlovitz numbers. In more applied configurations, Chung, Mishra, and Ihme developed interpretable SGS closures for transcritical LOX/GCH$_4$ mixing layers relevant to rocket engines and used feature-attribution analysis to relate the learned model behavior to physical mechanisms \cite{chung2022}. Physics-aware neural flame closures have also been used for combustion-instability modeling in a single-injector model engine \cite{shadram2022}. Together with the a posteriori LES--PDF studies discussed in Section~\ref{sec:chemacc} \cite{franke2017,readshaw2021,malik2021,ranade2019}, these works reinforce three key conclusions: coupled stability is more important than offline regression accuracy, bounded and physically constrained outputs improve robustness, and the computational cost of data generation and training must be considered together with inference cost when assessing the practical value of a learned closure.

\subsection{Filtered density function closures}
\label{sec:fdf}

The FDF methodology introduced in Section~\ref{sec:governing} treats the chemical source term exactly through Eq.~\eqref{eq:fdfquad}, while SGS convection and micro-mixing require modeling and the solution is carried by Lagrangian particles or Eulerian stochastic fields \cite{colucci1998,jaberi1999,givi2006}. The cost structure identifies three insertion points for learning. The first is particle chemistry, usually the dominant cost, which was in fact the earliest application of neural kinetics \cite{christo1996} and remains the natural deployment target for the NODE surrogates of Section~\ref{sec:chemacc}, since particle compositions evolve exactly as Eq.~\eqref{eq:kinetics} between mixing events \cite{franke2017,readshaw2021}. The second is the micro-mixing closure, whose key input, a mixing frequency proportional to the scalar dissipation rate, has been learned directly from filtered DNS \cite{yellapantula2021}, with conditional mixing statistics learned in reduced spaces \cite{ranade2019}. The third is the shape of the sub-filter distribution itself. Henry de Frahan et al. trained regression and generative models to predict the joint sub-filter PDF of mixture fraction and progress variable, outperforming $\beta$ presumptions particularly in multi-modal regions \cite{henrydefrahan2019}, and Chen et al. applied deep networks to the joint FDF in MILD combustion, a regime in which reaction zones are distributed and analytic joint distributions are inaccurate \cite{chen2021mild}.  Such learned distributions provide a data-driven approximation of the non-canonical sub-filter PDF shapes that transported FDF methods resolve more directly but at substantially higher computational cost. They also motivate hybrid FDF strategies in which a learned presumed distribution is used in regions with simple or weakly multi-modal composition structure, while Monte Carlo particles are retained only where the sub-filter composition distribution is strongly non-Gaussian, multi-modal, or otherwise difficult to represent with a low-cost closure.

Learning a distribution imposes stricter requirements than learning a scalar. A valid FDF must be non-negative and normalized,
\begin{equation}
F(\bm{\psi}) \geq 0, \qquad \int F(\bm{\psi})\,\mathrm{d}\bm{\psi} = 1,
\label{eq:fdfadmissible}
\end{equation}
its low-order moments must agree with the resolved or transported quantities that parameterize it, and its support must respect physical bounds: mixture fraction lies in $[0,1]$ and mass fractions are non-negative. If a learned FDF places probability outside the admissible composition space, the integrated source term of Eq.~\eqref{eq:fdfquad} becomes unphysical. These requirements can be enforced through output transformations, normalized basis expansions, mixture-density networks, or moment consistency penalties, in direct application of the soft and hard constraint strategies of Section~\ref{sec:embedding}.

\begin{figure}[tph]
\centering
\includegraphics[width=0.98\textwidth]{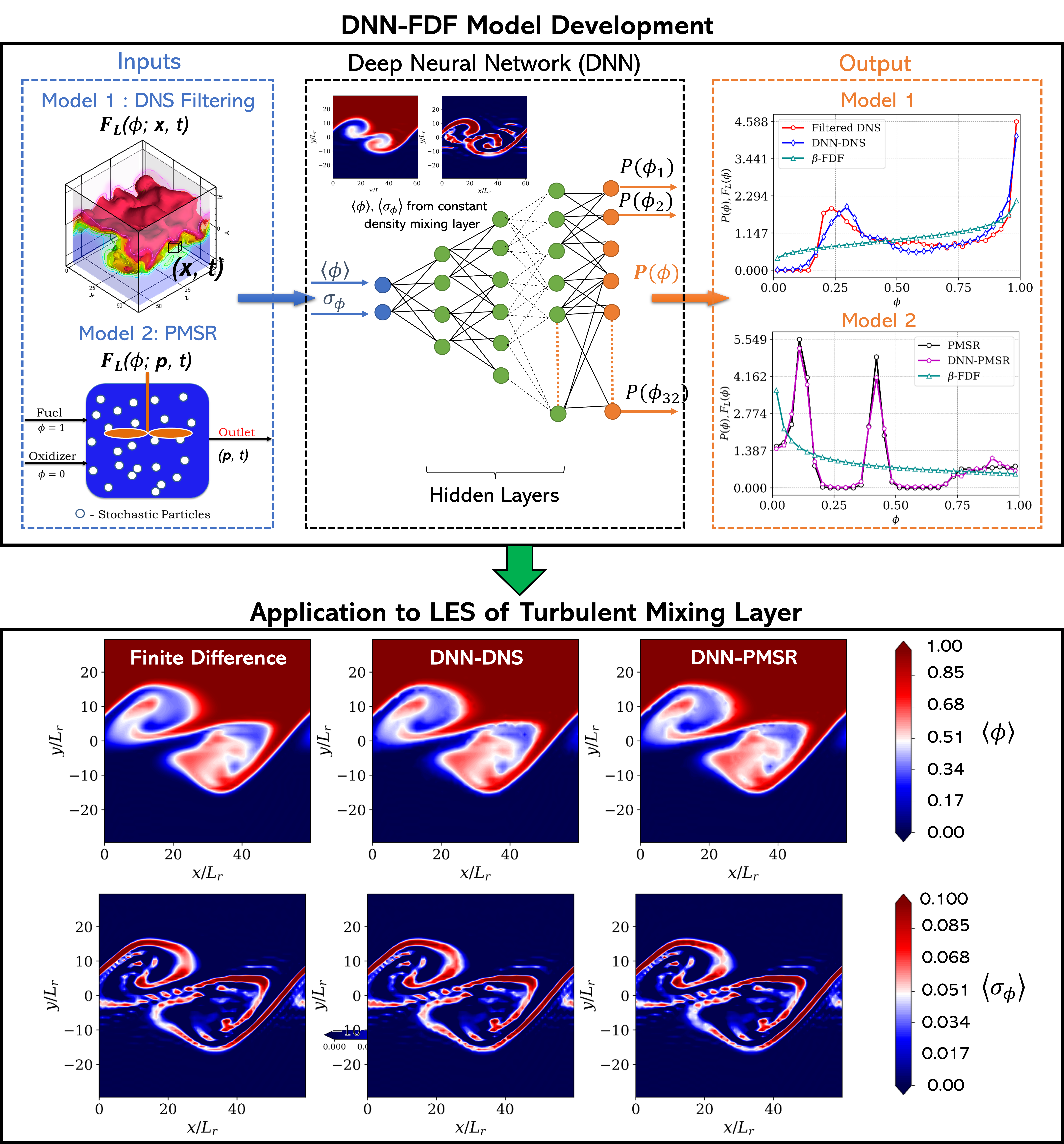}
\caption{Development and generalizable DNN-FDF model for LES of turbulent scalar mixing. The upper panel summarizes the DNN-FDF model development workflow, in which training FDFs are generated either from filtered DNS of constant density temporal mixing layers or from an ensemble of zero-dimensional PMSR simulations. The filtered mean and subgrid variance of mixture fraction,$\phi$, are used as inputs to the DNN, which predicts the corresponding sub-filter FDF. The lower panels show representative LES results for a temporal mixing layer at doubled grid resolution and half the original filter size, comparing the filtered mean mixture fraction and subgrid variance predicted using the DNN-FDF models with finite-difference reference results. The agreement demonstrates that the learned FDF closure generalizes across grid and filter sizes without retraining, provided the subgrid scale scalar variation is represented through the input variance \cite{bansude2024}.}
\label{fig:fdfresults}
\end{figure}

The author's DNN-FDF model \cite{bansude2024} addresses the FDF closure problem by learning the sub-filter distribution itself, rather than learning only selected filtered moments. The model is positioned between two established approaches: the transported FDF, which provides a detailed Monte Carlo representation of the subfilter scalar distribution but carries substantial particle cost \cite{jaberi1999,sheikhi2007}, and the presumed $\beta$-FDF, which is inexpensive but can become inaccurate when the local FDF is multi-modal, narrow, skewed, or otherwise inconsistent with the $\beta$ shape. In the DNN-FDF framework, the filtered mean and SGS variance of mixture fraction are used as physically motivated inputs, consistent with conventional presumed-FDF modeling, while a DNN predicts the full discretized FDF over the mixture fraction sample space. To ensure that the learned output remains an admissible probability distribution, a softmax activation function is used at the output layer, enforcing non-negative bin values that sum to unity; the resulting normalized output is then converted to the corresponding probability density over the mixture fraction bins. The predicted distribution is then used in Eq.~\eqref{eq:fdfquad} to evaluate filtered moments and filtered nonlinear functions. The model is a hybrid closure in the sense of Section~\ref{sec:embedding}: transport of the resolved fields remains conventional, and only the sub-filter distribution is learned.

The model was developed for the mixture fraction FDF in LES of variable density temporal mixing layers with conserved-scalar transport, following the validation hierarchy discussed in Section~\ref{sec:embedding}: the scalar mixing closure is first tested without finite-rate chemistry before being extended to reacting flow applications. A systematic procedure based on learning curves was used to select the training sample size and DNN architecture, thereby controlling bias and variance instead of choosing the model size ad hoc. Two variants were examined to test different data generation strategies. The DNN-DNS model was trained using FDFs obtained by filtering DNS of a constant density temporal mixing layer, while the DNN-PMSR model was trained using FDFs generated from a zero-dimensional pairwise mixing stirred reactor, providing a lower-cost alternative for situations in which DNS data are unavailable or prohibitively expensive. Both models reproduced the filtered scalar moments obtained from direct moment transport and showed improved behavior relative to the $\beta$-FDF, especially in high-variance and multi-modal regions where the assumed $\beta$ distribution is structurally limited. The summary of DNN-FDF model development is provided in  Fig.~\ref{fig:fdfresults}, upper panel. The models were also evaluated against transported FDF Monte Carlo simulations, demonstrating comparable accuracy for the first two filtered moments while avoiding sensitivity to the Monte Carlo ensemble domain size. Importantly, the study did not only assess the learned FDF shape; it also tested the ability of the predicted distribution to filter highly nonlinear scalar functions, which is directly relevant to chemical source term closure in reacting flows. The DNN-FDF models retained favorable accuracy in variable density mixing layers and showed generalization across different LES grid resolutions and filter sizes without retraining, because the subgrid scale variation was encoded through the input variance as shown in Fig.~\ref{fig:fdfresults}, bottom panel. These results show that learning the FDF distribution, when combined with physically motivated inputs, probability-preserving output structure, systematic data design, and comparison against both DNS and Monte Carlo FDF methods, provides a physics-guided alternative to conventional presumed-FDF closures.

\section{Cross-Cutting Requirements and Research Directions}
\label{sec:challenges}



The application areas discussed in Section~\ref{sec:pgml} share a common set of requirements that currently limit the transition of ML models from research demonstrations to routine engineering use. This section discusses these requirements and the research directions needed to address them.

\subsection{Generalization across fuels, combustion regimes, and configurations}
\label{sec:generalization}

Generalization remains one of the main barriers to the routine use of machine learning closures in turbulent combustion. A model trained on a single flame or operating condition is, in the strictest sense, a model of that specific dataset unless its inputs, architecture, and training distribution are designed to support transfer. The studies reviewed above suggest a hierarchy of transferability. Closures based on local, non-dimensional, and physically relevant inputs are generally more transferable. For example, SGS reaction-rate or scalar-dissipation closures expressed using quantities related to Karlovitz and Damk\"ohler numbers are more likely to generalize across filter sizes, grid resolutions, and moderate changes in operating conditions \cite{lapeyre2019,yellapantula2021}. Chemistry surrogates can transfer across flow configurations when the fuel, pressure range, and thermochemical manifold remain similar, provided that the training data adequately cover the composition states visited during deployment \cite{ding2021,wan2020}. Their transfer across fuels is more limited because the underlying chemical manifold changes. Configuration-level surrogates, such as models trained to reproduce an entire flame or combustor response, are typically the least transferable because they can encode geometry, boundary condition, and regime specific correlations.

Several strategies can improve generalization, although each introduces additional data, modeling, or computational requirements. Broader and multi-scale sampling can increase coverage of the relevant composition and parameter space \cite{zhang2022}. Embedded invariances and non-dimensional inputs can reduce the dependence of the model on coordinate systems, units, and specific flow scales \cite{ling2016}. Transfer learning can adapt a pretrained model to a new but related regime using a smaller amount of additional data. Operator learning formulations provide another promising direction because they are designed to learn mappings over families of initial conditions, parameters, or boundary conditions rather than a single trajectory or case \cite{goswami2024,weng2025}. Even with these strategies, extrapolation cannot be eliminated entirely. Therefore, extrapolation detection should be treated as an operational requirement rather than an optional diagnostic. A more rigorous assessment of generalization also requires shared benchmarks. Standardized filtered-DNS datasets, such as those being developed through initiatives like BLASTNet \cite{blastnet2022}, are important because they allow closures to be tested across common cases, filter widths, grid resolutions, and operating conditions. Such benchmarks would make it possible to measure generalization systematically rather than infer it from isolated studies. Looking further ahead, large heterogeneous datasets raise the possibility of pretrained representations of reacting flow fields that can be fine-tuned for specific closures using limited additional data. However, scale alone will not make such models reliable. The physics-guided principles discussed in Section~\ref{sec:embedding}, including physically meaningful inputs, conservation constraints, invariance, boundedness, uncertainty awareness, and solver-aware validation, should remain central to the development of these larger models.

\subsection{Interpretability}
\label{sec:interpretability}

Interpretability is important for two reasons. First, it allows learned combustion models to be checked against established physical understanding. Second, it can return useful modeling insight by revealing which variables, mechanisms, or interactions control the learned correction. Existing approaches span a spectrum from intrinsically interpretable models to post hoc diagnostic tools. At one end, CRNN embeds reaction mechanism structure directly into the architecture, so that the learned parameters can be interpreted in terms of stoichiometric coefficients, reaction orders, and rate constants \cite{ji2021crnn}. Sparse symbolic regression provides another interpretable strategy by identifying closed-form algebraic terms that can be examined, simplified, and compared with conventional closure assumptions \cite{schmelzer2020}. Constrained architectures such as TBNN also improve interpretability because the predicted quantities are expressed through invariant tensor bases, with learned scalar coefficients that depend on physically meaningful invariants \cite{ling2016}.

For more flexible models, interpretability is often introduced after training through feature attribution or sensitivity analysis. Chung et al.~\cite{chung2022}, for example, used feature importance analysis to determine which resolved flow quantities controlled learned SGS closures for transcritical combustion and connected these dependencies to the underlying physical mechanisms. Such analysis is especially valuable when a data-driven model outperforms a conventional closure. Identifying the inputs responsible for the improvement can reveal missing dependencies in the baseline model, such as sensitivity to local strain, scalar dissipation, density-gradient effects, heat release, or regime parameters. In this sense, interpretability should not be viewed only as a diagnostic for trust; it can also guide the development of improved physics-based closures. For physics-guided combustion ML, the most useful models are therefore not necessarily the most flexible black-box predictors, but those whose behavior can be interrogated, constrained, and related back to TCI physics \cite{ihme2022}.

\subsection{Uncertainty quantification}
\label{sec:uq}

A learned closure is difficult to trust if it does not provide some indication of its own reliability. This is especially important when the model is used outside the range of data on which it was trained, because neural networks can give confident but inaccurate predictions in such regions. In this context, it is useful to distinguish between two types of uncertainty \cite{kendall2017}. The first is \emph{aleatoric uncertainty}, which represents variability that is inherent in the data. In combustion modeling, this may arise because the same resolved inputs do not uniquely determine the exact unclosed term. In such cases, the model should ideally predict not only a mean value but also the spread of possible values. The second is \emph{epistemic uncertainty}, which represents uncertainty due to limited training data or insufficient model knowledge. This type of uncertainty can be reduced by adding more relevant data. More importantly, it usually increases when the model is queried far from the training distribution, making it useful for detecting out-of-distribution (OOD) states.

Several methods are available to estimate uncertainty in neural-network closures. Deep ensembles train multiple networks from different initial conditions and use the spread among their predictions as an estimate of model uncertainty \cite{lakshminarayanan2017}. Monte Carlo dropout provides a cheaper approximation by keeping dropout active during inference and evaluating the model multiple times \cite{gal2016}. Bayesian neural networks take a more formal approach by assigning probability distributions to the network weights, allowing uncertainty to be estimated as part of the model prediction \cite{kendall2017,geneva2019}. In reacting flow applications, Bayesian neural networks have been used to model the sub-filter progress variable dissipation rate and to identify regions where the available training data are insufficient \cite{pash2025}. For practical deployment, two points are especially important. First, uncertainty estimates must be calibrated, meaning that the predicted confidence should be consistent with the actual observed errors. For combustion, calibration should ideally be assessed using physically meaningful quantities such as ignition delay, extinction behavior, emissions, or instability margins. Second, the solver should be able to respond when uncertainty is too high. For example, it may switch back to a conventional closure or direct chemistry integration when the epistemic uncertainty exceeds a prescribed threshold. In this way, uncertainty quantification becomes not only a diagnostic tool but also a runtime safety mechanism. A major open challenge is uncertainty propagation: closure-level uncertainty must ultimately be carried through LES or RANS calculations to engineering quantities of interest, such as blow-off limits, pressure oscillations, and pollutant emissions \cite{duraisamy2019}.

\subsection{Data Infrastructure and Solver Integration}
\label{sec:data}

Progress in physics-guided combustion ML increasingly depends on reliable data infrastructure and robust solver integration, in addition to model architecture. Training datasets must be generated with consistent filtering operations, clearly documented thermochemical metadata, and well-defined input-output variables. Equally important is the design of validation splits. Randomly separating points from the same simulation can overestimate model performance; more meaningful tests should separate physical conditions such as flames, fuels, filter sizes, pressures, equivalence ratios, or operating regimes. Public DNS repositories and community datasets \cite{chen2011,blastnet2022} are therefore valuable because they provide common reference cases for evaluating and comparing learned closures. Since high-fidelity reacting flow data are expensive, future model development will likely rely on multi-fidelity datasets. Zero-dimensional reactors, laminar flames, RANS, LES, DNS, and experiments can provide complementary information at different levels of cost and fidelity. The PMSR-trained DNN-FDF model discussed in Section~\ref{sec:fdf} illustrates this idea: a lower-cost stochastic mixing configuration was used to generate training FDFs for situations where DNS data may be unavailable or impractical. Active learning can further improve data efficiency by identifying where additional simulations or experiments would be most useful, such as thermochemical states with high source term uncertainty, filter widths and variance levels associated with multi-modal FDFs, or combustor regions involving ignition, extinction, wall quenching, and strong heat release.

Deployment imposes a separate set of constraints. A learned closure must be evaluated repeatedly over many cells or particles, so inference cost, memory transfer, CPU/GPU placement, batching strategy, and compatibility with parallel solvers determine its practical benefit \cite{readshaw2021}. Robust implementations should also include fail-safe mechanisms. Inadmissible inputs, nonphysical outputs, or states far from the training distribution should trigger fallback to a conventional closure or detailed solver. In practical CFD calculations, this reliability requirement may be more important than maximizing average speed-up. Several research directions follow from these requirements. Differentiable reacting flow solvers could extend embedded-training ideas developed for momentum closures \cite{sirignano2020,macart2021} to stiff chemistry, FDF particle dynamics, scalar mixing, and multiphysics coupling, allowing closures to be trained based on their effect on the evolved solution rather than only on offline closure errors. For FDF methods, important targets include learned micro-mixing models that preserve boundedness and variance decay, realizable learned subfilter distributions, and adaptive hybrid presumed/transported schemes that retain transported particles only in strongly multi-modal regions. These developments are particularly relevant for hydrogen and ammonia combustion, where differential diffusion, thermodiffusive instabilities, wide flammability limits, and NO$_x$-critical chemistry challenge closures developed primarily for hydrocarbon flames. Across these directions, the physics-guided principles discussed in Section~\ref{sec:embedding} remain essential.

\section{Conclusions}
\label{sec:conclusion}

This article has reviewed the convergence of two modeling traditions. Conventional turbulent combustion closures encode substantial physical insight, but they remain constrained by chemical stiffness, empirical closure assumptions, and the dimensional growth of tabulated representations. Machine learning provides fast and flexible function approximation; however, without physical guidance, it can produce models that violate conservation, boundedness, thermochemical consistency, and numerical stability. Physics-guided machine learning provides the synthesis: physical knowledge is embedded in training data and input features, built into model architectures, incorporated through loss functions, structured through hybrid closures that learn only the uncertain components, and tested through solver-aware training and a posteriori validation.

The literature reviewed here shows this approach working across chemistry acceleration, reduced order modeling, subgrid closure, and FDF closure, with a posteriori demonstrations on the Sandia and Sydney flame series and on application-relevant configurations marking the transition from proof of concept toward practice. The author's studies, discussed within the corresponding topics, illustrate how selected elements of the framework operate in practice. The neural ODE kinetics model combines an architectural inductive bias with systematic composition-space sampling; the resulting learned dynamics are less stiff than the original mechanism, enabling explicit integration, and retain accuracy under the mixing-induced perturbations of the PMSR. The DNN-FDF closure combines a hybrid closure structure, which retains the exact source term treatment of the FDF formalism, with observational-bias strategies that include low-fidelity PMSR training data; it improves on the presumed $\beta$-FDF most where that assumption fails, for multi-modal and high-variance distributions. Neither study employs every approach in the taxonomy, and this selectivity reflects a general conclusion of the review: the appropriate combination of physics-guided strategies depends on the modeling target, the constraints whose violation would destabilize the solver, and the data that can realistically be generated.

Progress on generalization, coupled stability, uncertainty-aware deployment, interpretability, and solver integration will determine how quickly such models become standard components of predictive combustion simulation. The consistent lesson of the work surveyed here is that the physics should be treated not just as a constraint on machine learning but as its most informative prior.

\section*{Acknowledgement}

The author gratefully acknowledges the collaborators and mentors whose contributions and discussions informed the studies summarized in this article. 

\bibliographystyle{unsrt}
\bibliography{references}

\end{document}